\documentclass[12pt]{iopart}

\usepackage{graphicx}
\usepackage{rotating}

\usepackage{iopams}

\newtheorem{theorem}{Theorem}

\newtheorem{proposition}{Proposition}

\begin{document}

\title[]{Generalized conformal pseudo-Galilean algebras and their Casimir operators  }

\author{R. Campoamor-Stursberg\dag}
\address{\dag\ Instituto de Matem\'atica Interdisciplinar and Dpto. Geometr\'{\i}a y
Topolog\'{\i}a,\\
Universidad Complutense de Madrid, Plaza de Ciencias 3, E-28040
Madrid, Spain} \ead{rutwig@ucm.es}

\author{I. Marquette\ddag}
\address{\ddag School of Mathematics and Physics,\\ The University of Queensland, St Lucia QLD 4072, Australia}
\ead{i.marquette@uq.edu.au}

\begin{abstract}
A generalization $\mathfrak{Gal}_{\ell}(p,q)$ of the conformal Galilei algebra $\mathfrak{g}_{\ell}(d)$ with Levi subalgebra isomorphic to
$\mathfrak{sl}(2,\mathbb{R})\oplus\mathfrak{so}(p,q)$ is introduced and a virtual copy of the latter in the enveloping algebra of the extension 
is constructed. Explicit expressions for the Casimir operators are obtained from the determinant of polynomial matrices. For the central factor 
$\overline{\mathfrak{Gal}}_{\ell}(p,q)$, an exact formula giving the number of invariants is obtained and a procedure to compute invariants functions that 
do not depend on variables of the Levi subalgebra is developed. It is further shown that such solutions determine complete sets of invariants provided that the 
relation $d\leq 2\ell+2$ is satisfied.

\end{abstract}


\maketitle


\newpage

\section{Introduction}

Conformal invariance was first recognised to be of physical interest when it was realized that the Maxwell equations are covariant under the $15$-dimensional conformal group \cite{Cu,Bat}, a fact that motivated a more detailed analysis of conformal invariance in other physical contexts such as General Relativity, Quantum Mechanics or high energy physics \cite{Ful}. These applications further suggested to study conformal invariance in connection with the physically-relevant groups, among which the Poincar\'e and Galilei groups were the first to be considered. In this context, conformal extensions of the Galilei group have been considered in Galilei-invariant field theories, in the study of possible dynamics of interacting particles as well as in the nonrelativistic AdS/CFT correspondence
\cite{Bar54,Hag,Hav,Zak,Fig}. Special cases as the (centrally extended) Schr\"odinger algebra $\widehat{\mathcal{S}}(n)$ corresponding to the maximal invariance group of the 
free Schr\"odinger equation have been studied in detail by various authors, motivated by different applications such as the kinematical invariance of hierarchies of partial differential equations,  Appell systems, quantum groups or representation theory \cite{Ni72,Ni73,Do97,Fra}. The class of Schr\"odinger algebras can be generalized in natural manner to the so-called conformal Galilei algebras $\mathfrak{g}_{\ell}(d)$ for (half-integer) values $\ell\geq \frac{1}{2}$, 
also corresponding to semidirect products of the semisimple Lie algebra $\mathfrak{sl}(2,\mathbb{R})\oplus\mathfrak{so}(d)$ with a Heisenberg algebra but with a higher dimensional characteristic representation.\footnote{By characteristic representation we mean the representation of $\mathfrak{sl}(2,\mathbb{R})\oplus\mathfrak{so}(d)$  that describes the action on the Heisenberg algebra.} Such algebras, that can be interpreted as a nonrelativistic analogue of the conformal algebra, have been used in a variety of contexts, ranging from classical (nonrelativistic) mechanics, electrodynamics and fluid dynamics to higher-order Lagrangian mechanics \cite{Ai12,Tac,Du11,St13}
The algebraic structure of the conformal Galilei algebra $\mathfrak{g}_{\ell}(d)$ for values of $\ell\geq \frac{3}{2}$ and its representations have been analyzed in some detail, and an algorithmic procedures to compute their Casimir operators have been proposed (see e.g. \cite{Als17,Als19} and references therein). In the recent note \cite{raub}, a synthetic formula for the Casimir operators of the $\mathfrak{g}_{\ell}(d)$ algebra has been given. Although not cited explicitly, the  
procedure used there corresponds to the so-called ``virtual-copy" method, a technique well-known for some years that enables to compute the Casimir operators of a Lie algebra using those of its maximal semisimple subalgebra (\cite{Que,C23,C45,SL3} and references therein). 

\medskip
\noindent 
In this work, we first propose a further generalization of the conformal Galilei algebras $\mathfrak{g}_{\ell}(d)$, replacing the $\mathfrak{sl}(2,\mathbb{R})\oplus\mathfrak{so}(d)$  subalgebra of the latter by the semisimple Lie algebra $\mathfrak{sl}(2,\mathbb{R})\oplus\mathfrak{so}(p,q)$. As the defining representation $\rho_d$ of $\mathfrak{so}(p,q)$ is real for all values $p+q=d$ \cite{Tits}, the structure of a semidirect product with a Heisenberg Lie algebra remains unaltered. The Lie algebras $\mathfrak{Gal}_{\ell}(p,q)$ describe a class of semidirect products of semisimple and Heisenberg Lie algebras among which $\mathfrak{g}_{\ell}(d)$ corresponds to the case with a largest maximal compact subalgebra. 
Using the method developed in \cite{C45}, we construct a virtual copy of $\mathfrak{sl}(2,\mathbb{R})\oplus\mathfrak{so}(p,q)$ in the enveloping algebra of $\mathfrak{Gal}_{\ell}(p,q)$ for all half-integer values of $\ell$ and any $d=p+q\geq 3$. The Casimir operators of these Lie algebras are determined combining the analytical and the matrix trace methods, showing how to compute them explicitly in terms of the determinant of a polynomial matrix.

\medskip
\noindent We further determine the exact number of Casimir operators for the unextended Lie algebras $\overline{\mathfrak{Gal}}_{\ell}(p,q)$ obtained by factorizing 
$\mathfrak{Gal}_{\ell}(p,q)$ by its centre. Using the reformulation of the Beltrametti-Blasi formula in terms of the Maurer-Cartan equations, we show that albeit the number $\mathcal{N}$ of invariants increases considerably for fixed $\ell$ and varying $d$, a generic polynomial formula at most quadratic in $\ell$ and $d$ that gives the exact value of $\mathcal{N}$ can be established. Depending on the fact whether the relation $d\leq 2\ell+2$ is satisfied or not, it is shown that $\overline{\mathfrak{Gal}}_{\ell}(p,q)$ admits a complete set of invariants formed by operators that do not depend on the generators of the Levi subalgebra. An algorithmic procedure to compute these invariants by means of a reduction to a linear system is proposed.

\section{Maurer-Cartan equations of Lie algebras and Casimir operators  }

Given a Lie algebra $ \frak{g}=\left\{X_{1},..,X_{n}\; |\;
\left[X_{i},X_{j}\right]=C_{ij}^{k}X_{k}\right\}$ in terms of
generators and commutation relations, we are principally interested
on (polynomial) operators
$C_{p}=\alpha^{i_{1}..i_{p}}X_{i_{1}}..X_{i_{p}}$ in the
generators of $\frak{s}$ such that the constraint $
\left[X_{i},C_{p}\right]=0$,\; ($i=1,..,n$) is satisfied. Such an
operator can be shown to lie in the centre of the enveloping
algebra of $\frak{g}$ and is called a (generalized) Casimir
operator. For semisimple Lie algebras, the determination of
Casimir operators can be done using structural properties
\cite{Ra,Gel}. However, for non-semisimple Lie algebras the relevant
invariant functions are often rational or even transcendental
functions \cite{Bo1,Bo2}. This suggests to develop a method in order to
cover arbitrary Lie algebras. One convenient approach is the
analytical realization. The generators of the Lie algebra
$\frak{s}$ are realized in the space $C^{\infty }\left(
\frak{g}^{\ast }\right) $ by means of the differential operators:
\begin{equation}
\widehat{X}_{i}=C_{ij}^{k}x_{k}\frac{\partial }{\partial x_{j}},
\label{Rep1}
\end{equation}
where $\left\{ x_{1},..,x_{n}\right\}$ are the coordinates  in a dual basis of
$\left\{X_{1},..,X_{n}\right\} $. The invariants of $\frak{g}$ hence correspond to solutions of the following
system of partial differential equations:
\begin{equation}
\widehat{X}_{i}F=0,\quad 1\leq i\leq n.  \label{sys}
\end{equation}
Whenever we have a polynomial solution of (\ref{sys}), the
symmetrization map defined by
\begin{equation}
{\rm Sym(}x_{i_{1}}^{a_{1}}..x_{i_{p}}^{a_{p}})=\frac{1}{p!}\sum_{\sigma\in
S_{p}}x_{\sigma(i_{1})}^{a_{1}}..x_{\sigma(i_{p})}^{a_{p}}\label{syma}
\end{equation}
allows to rewrite the Casimir operators in their usual form 
as central elements in the enveloping algebra of $\frak{g}$,
after replacing the variables $x_{i}$ by the corresponding
generator $X_{i}$. A maximal set of functionally
independent invariants is usually called a fundamental basis. The
number $\mathcal{N}(\frak{g})$ of functionally independent
solutions of (\ref{sys}) is obtained from the classical criteria
for differential equations, and is given by the formula  
\begin{equation}
\mathcal{N}(\frak{g}):=\dim \,\frak{g}- {\rm
sup}_{x_{1},..,x_{n}}{\rm rank}\left( C_{ij}^{k}x_{k}\right),
\label{BB}
\end{equation}
where $A(\frak{g}):=\left(C_{ij}^{k}x_{k}\right)$ is the matrix
associated to the commutator table of $\frak{g}$ over the given
basis \cite{Be}.\newline 
The reformulation of condition (\ref{BB}) in terms of differential forms (see e.g. \cite{C43})
allows to compute  $\mathcal{N}(\frak{g})$ quite efficiently and even to 
obtain the Casimir
operators under special circumstances \cite{Peci,C72}. In terms of the
Maurer-Cartan equations, the Lie algebra $\frak{g}$
is described as follows: If $\left\{  C_{ij}
^{k}\right\}  $ denotes the structure tensor over the basis $\left\{  X_{1},..,X_{n}\right\} $,
the identification of the dual space $\frak{g}^{\ast}$ with the
left-invariant 1-forms on the simply connected Lie group the Lie algebra of which is isomorphic to $\frak{g}$ allows to define an exterior
differential $d$ on $\frak{g}^{\ast}$ by
\begin{equation}
d\omega\left(  X_{i},X_{j}\right)  =-C_{ij}^{k}\omega\left(
X_{k}\right) ,\;\omega\in\frak{g}^{\ast}.\label{MCG}
\end{equation}
Using the coboundary operator $d$, we rewrite $\frak{g}$ as a
closed system of $2$-forms%
\begin{equation}
d\omega_{k}=-C_{ij}^{k}\omega_{i}\wedge\omega_{j},\;1\leq
i<j\leq\dim\left( \frak{g}\right)  ,\label{MC2}
\end{equation}
called the Maurer-Cartan equations of $\frak{g}$. 
In order to reformulate equation (\ref{BB}) in this context, we consider the linear subspace
$\mathcal{L}(\frak{g})=\mathbb{R}\left\{ d\omega_{i}\right\}
_{1\leq i\leq \dim\frak{g}}$ of $\bigwedge^{2}\frak{g}^{\ast}$
generated by the $2$-forms $d\omega_{i}$. Now, for 
a generic element  $\omega=a^{i}d\omega_{i}\,\;\left(
a^{i}\in\mathbb{R}\right)  $ of $\mathcal{L}(\frak{g})$ there
exists a positive integer $j_{0}\left( \omega\right)
\in\mathbb{N}$ such that $\bigwedge^{j_{0}\left( \omega\right)
}\omega\neq0$ and $\bigwedge ^{j_{0}\left( \omega\right)
+1}\omega\equiv0$. We define the scalar $j_{0}\left(
\frak{g}\right) $ as the maximal rank of generic elements,
\begin{equation}
j_{0}\left(  \frak{g}\right)  =\max\left\{  j_{0}\left(
\omega\right) \;|\;\omega\in\mathcal{L}(\frak{g})\right\},
\label{MCa1}
\end{equation}
As shown in \cite{C43}, this is a scalar invariant of the Lie algebra $\frak{g}$ that
satisfies the relation
\begin{equation}
\mathcal{N}\left(  \frak{g}\right)  =\dim\frak{g}-2j_{0}\left(  \frak{g}%
\right). \label{BB1}
\end{equation}

\medskip

\section{Virtual copies of semisimple Lie algebras}

\noindent The method of virtual copies, essentially developed in \cite{SL3}, constitutes a natural generalization 
of a method due to Ch. Quesne (see \cite{Que}) that combines the boson formalism and enveloping algebras of Lie algebras in 
order to compute Casimir operators  of semidirect products
$\frak{s}\overrightarrow {\frak{\oplus}}_{R}\frak{r}$ of simple Lie algebras $\frak{s}$ and solvable algebras $\mathfrak{r}$.

\medskip
\noindent We briefly recall the procedure, the details of which can be found in \cite{SL3}: Let  $\frak{g}$ be 
a non-semisimple Lie algebra admitting the Levi decomposition
$\frak{g}=\frak{s}\overrightarrow{\frak{\oplus}}_{\Gamma}\frak{r}$,
where $\frak{s}$ denotes the Levi subalgebra, $\Gamma$ the characteristic representation and $\frak{r}$ the
radical, i.e., the maximal solvable
ideal of $\frak{g}$. Let $\left\{
X_{1},..,X_{n},Y_{1},..,Y_{m}\right\} $ be a basis such that
$\left\{  X_{1},..,X_{n}\right\}  $ spans $\frak{s}$ and $\left\{
Y_{1},..,Y_{m}\right\}  $ spans $\frak{r}$. We further suppose
that the structure tensor in $\frak{s}$ is given by
\begin{equation}
\left[  X_{i},X_{j}\right]  =C_{ij}^{k}X_{k}.\label{ST}
\end{equation}
We now define operators $X_{i}^{\prime}$ in the enveloping algebra
of $\frak{g}$ by means of
\begin{equation}
X_{i}^{\prime}=X_{i}\,f\left(  Y_{1},..,Y_{m}\right) +P_{i}\left(
Y_{1},..,Y_{m}\right)  ,\label{OP1}%
\end{equation}
where $P_{i}$ is a homogeneous polynomial of degree $k$ and $f$ is
homogeneous of degree $k-1$. We require
the constraints
\begin{eqnarray}
\left[  X_{i}^{\prime},Y_{k}\right]    & =0,\label{Bed1}\\
\left[  X_{i}^{\prime},X_{j}\right]    & =\left[
X_{i},X_{j}\right] ^{\prime}:=C_{ij}^{k}\left(
X_{k}f+P_{k}\right).\label{Bed2}
\end{eqnarray}
to be satisfied for all generators. This leads to 
conditions on $f$ and $P_{i}$.   It can be shown that condition (\ref{Bed1}) leads to
\begin{equation}
\left[  X_{i}^{\prime},Y_{j}\right]  =\left[  X_{i}f,Y_{j}\right]  +\left[  P_{i}%
,Y_{j}\right]  =X_{i}\left[  f,Y_{j}\right]  +\left[
X_{i},Y_{j}\right]
\,f+\left[  P_{i},Y_{j}\right]  .\label{Eq1}%
\end{equation}
By homogeneity, we can reorder the terms according to
their degree, so that $X_{i}\left[  f,Y_{j}\right]  $ is 
homogeneous of degree $k-1$ in the variables $\left\{
Y_{1},..,Y_{m}\right\}  $ and  $\left[ X_{i},Y_{j}\right]
\,f+\left[  P_{i},Y_{j}\right]  $ of degree $k$. Hence the conditions 
\begin{eqnarray}
\left[  f,Y_{j}\right] =0,\;
\left[  X_{i},Y_{j}\right]  \,f+\left[  P_{i},Y_{j}\right]
=0\label{Eq1A}
\end{eqnarray}
are satisfied, showing that $f$ is a Casimir operator
of the radical $\frak{r}$. Expanding the condition (\ref{Bed2}) and taking into 
account the homogeneity degrees, after a routine computation we find that the system
\begin{eqnarray}
\left[  X_{i},X_{j}\right]  \,f-X_{i}\left[  X_{j},f\right]    =C_{ij}
^{k}X_{k}f,\quad 
\left[  P_{i},X_{j}\right]    =C_{ij}^{k}P_{k}\label{Eq3}
\end{eqnarray}
is satisfied for any indices $i,j$. Using now 
(\ref{ST}), the first identity reduces to
\begin{equation}
X_{i}\left[  X_{j},f\right]  =0.
\end{equation}
From this we conclude that the function $f$ is a Casimir operator of $\frak{g}$ that depends
only on the variables of the radical $\frak{r}$. The second
identity in (\ref{Eq3}) implies that $P_{i}$ transforms under the
$X_{j}^{\prime}s$ like a generator of the semisimple part
$\frak{s}$. Taken together, it follows that the operators
$X_{i}^{\prime}$ fulfill the condition 
\begin{eqnarray}
\left[  X_{i}^{\prime},X_{j}^{\prime}\right]    & =f\left[  X_{i},X_{j}\right]  ^{\prime}.
\end{eqnarray}
We shall say that the operators $X_{i}^{\prime}$
generate a virtual copy of $\frak{s}$ in the enveloping algebra of
$\frak{g}$. If $f$ can be
identified with a central element of $\mathfrak{g}$, as happens for a radical isomorphic to a Heisenberg
algebra, the virtual copy actually generates a copy in
$\mathcal{U}\left(  \frak{g}\right) $ \cite{Que,C23}.  The computation of the invariants of $\mathfrak{g}$ reduces to application of the following result proved in \cite{SL3}:

\begin{theorem}
Let $\frak{s}$ be the Levi subalgebra of $\frak{g}$ and
let $X_{i}^{\prime}=X_{i}\,f\left(  \mathbf{Y}\right)
+P_{i}\left(\mathbf{Y}\right)  $ be homogeneous polynomials in the
generators of $\frak{g}$ satisfying equations (\ref{Eq1A}) and
(\ref{Eq3}). If $C=\sum\alpha ^{i_{1}..i_{p}}X_{i_{1}}..X_{i_{p}}$
is a Casimir operator of $\frak{s}$ having degree $p$, then
$C^{\prime}=\sum\alpha^{i_{1}..i_{p}}X_{i_{1}}^{\prime
}..X_{i_{p}}^{\prime}$ is a Casimir operator of $\frak{g}$ of
degree $(\deg f+1)p$. In particular, $\mathcal{N}\left(  \frak{g}\right)
\geq\mathcal{N}\left( \frak{s}\right)  +1.$
\end{theorem}

\medskip
\noindent 
The independence of the invariants obtained in such manner follows at once from the
conditions (\ref{Bed1}) and (\ref{Bed2}). For the particular case of a radical isomorphic to a 
Heisenberg Lie algebra, it follows that the number of non-central invariants is given by the rank 
of the semisimple part, i.e., $\mathcal{N}\left(  \frak{g}\right)
=\mathcal{N}\left( \frak{s}\right)  +1$ (see \cite{C45} for a proof).

\section{The conformal generalized pseudo-Galilean Lie algebra $\mathfrak{Gal}_{\ell}(p,q)$ }

\noindent Structurally, the conformal Galilean algebra  $\mathfrak{\widehat{g}}_\ell(d)$ is a semidirect product of the semisimple Lie algebra $\mathfrak{s}=\mathfrak{sl}(2,\mathbb{R})\oplus \mathfrak{so}(d)$ and a Heisenberg Lie algebra of dimension $N= d(2\ell+1)+1$. The action of $\mathfrak{s}$ over the radical is given by the characteristic representation 
$\widehat{\Gamma}=\left(D_{\ell}\otimes \rho_d\right)\oplus \Gamma_0$, where  $D_{\ell}$ denotes the irreducible representation of $\mathfrak{sl}(2,\mathbb{R})$ with highest weight $2\ell$ and dimension $2\ell+1$, $rho_d$ is the defining $d$-dimensional representation of $\mathfrak{so}(d)$ and $\Gamma_0$ denotes the trivial representation. 

\noindent 
Considering the basis (see e.g \cite{Als17}) given by the generators  $\left\{ H, D, C, E_{ij}=-E_{ji}, P_{n,i}\right\}$
with $n = 0,1, 2, \ldots, 2\ell; \, i,j =1, 2, \ldots,  d$, 
the commutators are   
\begin{eqnarray}
\fl [D,H]= 2H,\quad [D,C]= -2C,\quad [C, H]=D,  \nonumber\\
\fl[H,P_{n,i}]=-nP_ {n-1,i},\,\,[D,P_{n,i}]=2(\ell - n)P_ {n,i}, \,\,[C,P_{n,i}]=(2\ell - n)P_{n+1,i},\nonumber \\
\fl[E_{ij}, P_{n,k} ]=\delta_{ik}P_{n,j} - \delta_{jk}P_{n,i},  \,\,
[E_{ij}, E_{k\ell}] =\delta_{ik}E_{j\ell} + \delta_{j \ell}E_{ik} - \delta_{i \ell}E_{jk} - \delta_{jk}E_{i \ell} \nonumber\\
\fl \left[ P_{m,i},P_{n,j}\right]= \delta _{ij}  \delta _{m+n,2\ell} I_{m}  M ,\quad\qquad I_{m}= (-1)^{m+\ell+1/2}  (2\ell -m)! \,\ m! \label{CG3}
\end{eqnarray}

\noindent The invariants can be deduced from the Casimir operators of the semisimple subalgebra of $\mathfrak{g}$ by replacing the generators by expressions of the type 
(\ref{OP1}) that generate a virtual copy of $\mathfrak{s}$. For the case of the conformal generalized Galilean algebra $\widehat{\mathfrak{g}}_{\ell}(d)$, these invariants have recently been given implicitly in \cite{raub} essentially applying this method, although the nomenclature use there is referred to as ``disentaglement" of the generators.

\subsection{The conformal generalized pseudo-Galilean algebra}

\noindent Introducing a non-degenerate metric tensor of signature $(p,q)$, the structure of conformal Galilean algebras can be easily extended to the pseudo-orthogonal Lie algebras $\mathfrak{so}(p,q)$ ($p+q=d$) along the same lines. 
The pseudo-orthogonal algebra $\frak{so}(p,q)$ with
$d=p+q$ is given by the  $\frac{1}{2}d(d-1)$ operators
$E_{\mu\nu}=-E_{\nu\mu}$   satisfying:
\begin{eqnarray*}
\left[ E_{\mu \nu },E_{\lambda \sigma }\right]  &=&g_{\mu \lambda
}E_{\nu \sigma }+g_{\mu \sigma }E_{\lambda \nu }-g_{\nu \lambda
}E_{\mu \sigma
}-g_{\nu \sigma }E_{\lambda \mu } \\
\left[ E_{\mu \nu },P_{\rho }\right]  &=&g_{\mu \rho }P_{\nu
}-g_{\nu \rho }P_{\mu },
\end{eqnarray*}
where $g={\rm diag}\left( 1,..,1,-1,..,-1\right)$ is the matrix of the non-degenerate metric. Let $\rho_1$ be the $d$-dimensional defining representation  of $\frak{so}(p,q)$ and define the tensor product $\Gamma=D_{\ell}\otimes \rho_1$ for $\ell\in\mathbb{Z}+\frac{1}{2}$. Then $\Gamma$ is an irreducible representation of the semisimple Lie algebra $\mathfrak{sl}(2,\mathbb{R})\oplus \mathfrak{so}(p,q)$ that satisfies the condition $\Gamma_0\subset \Gamma\wedge \Gamma $, i.e., the wedge product of $\Gamma$ contains a copy of the trivial representation. Following the characterization given in \cite{C45}, this implies that the Lie algebra $\left(\mathfrak{sl}(2,\mathbb{R})\oplus \mathfrak{so}(d)\right)\overrightarrow{\oplus}_{\Gamma\oplus\Gamma_0}\mathfrak{h}_{N}$
with $N=d(2\ell+1)$ is well defined. Over the basis ${ H, D, C, E_{ij}=-E_{ji}, P_{n,i},M}$ with  $0\leq n \leq 2\ell$  and $1\leq i<j\leq p+q$, the brackets  are given by 
\begin{eqnarray}
[D,H]= 2H,\quad [D,C]= -2C,\quad [C, H]=D,  \nonumber\\
\,\,[H,P_{n,i}]=-nP_ {n-1,i},\,\,[D,P_{n,i}]=2(\ell - n)P_ {n,i}, \,\,[C,P_{n,i}]=(2\ell - n)P_{n+1,i},\label{CG3} \\
\,\,[E_{ij}, P_{n,k} ]=g_{ik}P_{n,j} - g_{jk}P_{n,i},  \,\,
[E_{ij}, E_{k\ell}] =g_{ik}E_{j\ell} + g_{j \ell}E_{ik} - g_{i \ell}E_{jk} - g_{jk}E_{i \ell}, \nonumber\\
\left[ P_{n,k},P_{m,l}\right]= g_{ij}  \delta _{m+n,2\ell} I_{m}  M ,\quad\qquad I_{m}= (-1)^{m+\ell+1/2}  (2\ell -m)! \,\ m!.\nonumber
\end{eqnarray}

\noindent As commented above, the number of Casimir operators is given by $2+\left[\frac{d}{2}\right]$ and can be deduced in closed form by 
means of the virtual copy method. 

\begin{proposition}
For any $\ell\in\mathbb{Z}+\frac{1}{2}\geq \frac{1}{2}$, the operators%
\begin{eqnarray}
\fl \widetilde{D}=D\,M+\sum_{i=1}^{d}\sum_{s=0}^{q}\left( -1\right) ^{s+q-1}\frac{\mu
^{1}\left( s,q\right)}{g_{ii}} P_{s,i}P_{2l-s,i},\nonumber \\
\fl \widetilde{H}=H\,M+\sum_{i=1}^{d}\sum_{s=0}^{q-1}\left( -1\right) ^{s+q-1}\frac{\mu
^{2}\left( s,q\right)}{g_{ii}} P_{s,i}P_{2l-1-s,i}-\sum_{i=1}^{d}\frac{1}{2\Gamma(q+1)^2g_{ii}}   P_{q,i}^2,\nonumber \\
\fl\widetilde{C}=C\,M+\sum_{i=1}^{d}\sum_{s=0}^{q}\left( -1\right) ^{s+q}\frac{\mu
^{3}\left( s,q\right)}{g_{ii}} P_{s,i}P_{2l+1-s,i}-\sum_{i=1}^{d}\frac{1}{2\Gamma(q+1)^2g_{ii}}   P_{q+1,i}^2,\nonumber \\
\fl \widetilde{E}_{i,j}=M E_{i,j}+ \sum_{s=0}^{l}\frac{(-1)^{\frac{2l-1}{2}+s}}{s!\; (2l-s)!}\left(P_{s,i}P_{2l-s,j}-P_{s,j}P_{sl-s,i}\right),\; 1\leq i<j\leq d,
\label{NE3}
\end{eqnarray}
with coefficients defined by 
\begin{eqnarray}
\fl \mu ^{1}\left( s,q\right) =2^{\frac{s-2}{2}}\left( 1+\sqrt{2}+\left( -1\right)
^{s}\left( \sqrt{2}-1\right) \right) \prod_{a=0}^{\left[ \frac{s+1}{2}\right]
-1}\left( q-\left[ \frac{s}{2}\right] -a\right) \prod_{b=s+1-\left[ \frac{s}{%
2}\right] }^{s}\left( 2q+3-2b\right) ,\nonumber\\
\fl \mu ^{2}\left( s,q\right) =\frac{1}{s!\; \Gamma(2q+1-s)},\quad \mu ^{3}\left( s,q\right) =\frac{1}{(s-1)!\; \Gamma(2q+2-s)}
\end{eqnarray}
generate a (virtual) copy of $\mathfrak{sl}(2,\mathbb{R})\oplus\frak{so}\left(p,q\right)  $ in the
enveloping algebra of $\mathfrak{Gal}_{\ell}(p,q)$.
\end{proposition}

\noindent The proof, albeit long and computationally cumbersome, is completely straightforward  and reduces to a direct verification of the
conditions (\ref{Bed1}) and (\ref{Bed2}) with the choice $f=M$, taking into account the following relations between the generators and quadratic products: 
\begin{eqnarray*}
 \left[D,P_{n,i}P_{m,j}\right]=2\left(2\ell-m-n\right)\;P_{n,i}P_{m,j},\\
 \left[H,P_{n,i}P_{m,j}\right]=-\left(n P_{n-1,i}P_{m,j}+m P_{n,i}P_{m-1,j}\right),\\
 \left[D,P_{n,i}P_{m,j}\right]=(2\ell-m)P_{n+1,i}P_{m,j}+(2\ell-m)M P_{n,i}P_{m+1,j},\\
 \left[M E_{i,j},M E_{k,l}\right]=M^2\left(g_{i,k}E_{j,l}+g_{j,l}E_{i,k}-g_{i,l}E_{j,k}-g_{j,k}E_{i,l}\right),\\
 \left[E_{i,j},P_{n,k}P_{m,l}\right]=-\left(g_{i,k}P_{n,j}P_{m,l}-g_{j,l}P_{n,k}P_{m,i}+g_{i,l}P_{n,k}P_{m,j}-g_{j,k}P_{n,i}P_{m,l}\right) ,\\
\left[P_{n,i}P_{m,j},P_{q,k}\right]=-I_{q}M\left(g_{i,k}\delta_{n+q}^{2\ell}P_{m,j}+g_{j,k}\delta_{m+q}^{2\ell}P_{n,i}\right).\\
\end{eqnarray*}
In particular, for the metric tensor $g_{ii}=1$ corresponding to the compact orthogonal algebra $\mathfrak{so}(d)$, we obtain an equivalent realization to 
the disentaglement conditions given in \cite{raub}.

\subsection{Explicit formulae for the Casimir operators of $\mathfrak{Gal}_{\ell}(p,q)$} 

\noindent Once the (virtual) copy of the semisimple Lie algebra $\mathfrak{Gal}_{\ell}(p,q)$ is found, explicit expression for the Casimir operators can be immediately deduced, in its 
unsymmetrized analytic form, by means of the well known trace methods (see e.g. \cite{Ra,Gel,Gr64,Per,Po66,Ok77,Mac}). To this extent, let $\left\{d,h,c,e_{i,j},p_{n,k}\right\}$ 
be the coordinates in $\mathfrak{Gal}_{\ell}(p,q)^{\ast}$ and let $\left\{\widehat{d},\widehat{h},\widehat{c},\widehat{e}_{i,j},\widehat{p}_{n,k}\right\}$ denote the analytical counterpart of the operators in (\ref{NE3}). As the simple subalgebras  $\mathfrak{sl}(2,\mathbb{R})$ and $\mathfrak{so}(p,q)$ commute, it follows at once that any invariant of $\mathfrak{Gal}_{\ell}(p,q)$ must be also an invariant of the subalgebra $\mathfrak{sl}(2,\mathbb{R})\overrightarrow{\oplus}_{\Gamma}\mathfrak{h}_N$. Semidirect products of $\mathfrak{sl}(2,\mathbb{R})$ and a Heisenberg
Lie algebra are well-known to possess only one Casimir operators besides the central generator \cite{C45}, the analytic expression of which is given by 
\begin{equation}
C^{\prime}_{4}= \widehat{d}^2-4\widehat{c}\widehat{h}\label{ins}
\end{equation}
This invariant can also be described as a determinant as follows (see e.g. \cite{C23}): Let $B=\left\{X_{2\ell(k-1)+k+2+s}\:\; 1\leq k\leq d,\; 1\leq s\leq 2\ell+1\right\}$ be a basis of 
$\left(\mathfrak{sl}(2,\mathbb{R})\overrightarrow{\oplus}_{\Gamma}\mathfrak{h}_N\right)$ such that $\left\{D,H,C\right\}=\left\{X_1,X_2,X_3\right\}$ and such that the element 
$X_{2\ell(k-1)+k+2+s}$ coresponds to the generator $P_{s,k}$ for $1\leq k\leq d,\; 1\leq s\leq 2\ell+1$. The commutators of $\left(\mathfrak{sl}(2,\mathbb{R})\overrightarrow{\oplus}_{\Gamma}\mathfrak{h}_N\right)$ are then described in uniform manner by 
\begin{equation}
\left[X_i,X_j\right]= C_{ij}^{k}X_{k},\; 1\leq i<j,k\leq (d+1)(2\ell+1).
\end{equation}
Let $B^{\ast}=\left\{x_{2\ell(k-1)+k+2+s}\:\; 1\leq k\leq d,\; 1\leq s\leq 2\ell+1\right\}$ be the dual basis of $B$ and define be the polynomial matrix  $A$  of order $4 + (2 \ell + 1) d$, the entries of which are given by 
\begin{eqnarray}
A_{i,j}= C_{ij}^{k} x_{k},\quad 1\leq i,j\leq 3 + (2 \ell + 1) d,\nonumber\\
A_{i,2 + (2 \ell + 1) d}= -A_{2 + (2 \ell + 1) d,i}x_{i},\quad 1\leq i\leq 3,\label{maxa}\\
A_{j,2 + (2 \ell + 1) d}=-A_{2 + (2 \ell + 1) d,j}=\frac{1}{2}x_j,\quad j\geq 4.\nonumber
\end{eqnarray} 
It follows from the analysis in \cite{C23} that the determinant $\det{A}$ provides the non-central Casimir invariant of  the Lie algebra $\left(\mathfrak{sl}(2,\mathbb{R})\overrightarrow{\oplus}_{\Gamma}\mathfrak{h}_N\right)$ . Comparing the result with that deduced from (\ref{ins}) using the copy in the enveloping algebra, we have the relation  
\begin{equation}
\det(A)=\prod_{s=1}^{d}\prod_{m=0}^{2\ell}\left(2\ell-m\right)!m!\;M^{2\ell d+d-4}\left(C_{4}^{\prime}\right)^2.\label{insa}
\end{equation}
 
\medskip
\noindent Similarly, we can consider the invariants of $\mathfrak{Gal}_{\ell}(p,q)$ that are simultaneously invariants of the subalgebra $\mathfrak{so}(p,q)\overrightarrow{\oplus}_{\Gamma}\mathfrak{h}_N$ with $d=p+q$.  
For the pseudo-orthogonal Lie algebra $\frak{so}(p,q)$, a maximal set of Casimir operators is well known to be given
by the coefficients $C_{k}$ of the characteristic polynomial $P(T)$ of the matrix 
\begin{equation}
B_{p,q}:=\left(
\begin{array}{cccccc}
0 & .. & -g_{jj}e_{1j} & .. & -g_{NN}e_{1,N}  \\
: &  & : &  & : &  \\
g_{11}e_{1j} & .. & 0 & .. & -g_{NN}e_{j,N}  \\
: &  & : &  & : &  \\
g_{11}e_{1,N} & .. & g_{jj}e_{j,N} & .. & 0 & 
\end{array}
\right)\label{MA2}
\end{equation}
 
\noindent The same formula, replacing the generators $e_{i,j}$ by those $\widetilde{e}_{i,j}$ of the virtual copy will provide us with  the invariants of $\mathfrak{Gal}_{\ell}(p,q)$ that only depend on the generators of $\frak{so}(p,q)$ and the characteristic representation $\Gamma$.  

\begin{proposition}
A maximal set of $\left[\frac{d}{2}\right]$ independent Casimir operators of $\mathfrak{Gal}_{\ell}(p,q)$ depending only on the generators of  $\frak{so}(p,q)$ and the $\left\{P_{0,i},\cdots P_{2\ell,i}\right\}$ with $1\leq i\leq p+q=d$  
is given by the coefficients $\widetilde{C}_{k}$ of the polynomial $P(T)$ defined by
\begin{equation}
P(T):=\det \left( B_{p,q}-T\;\mathrm{Id}_{N}\right)  ,  \label{Pol1}
\end{equation}
where
\begin{equation}
B_{p,q}:=\left(
\begin{array}{cccccc}
0 & .. & -g_{jj}\widetilde{e}_{1j} & .. & -g_{NN}\widetilde{e}_{1,N}  \\
: &  & : &  & : &  \\
g_{11}\widetilde{e}_{1j} & .. & 0 & .. & -g_{NN}\widetilde{e}_{j,N}  \\
: &  & : &  & : &  \\
g_{11}\widetilde{e}_{1,N} & .. & g_{jj}\widetilde{e}_{j,N} & .. & 0 & 
\end{array}
\right)
\end{equation}
\end{proposition}
The actual symmetric representatives ${\rm Sym}(\widetilde{C}_k)$ of the invariants as elements in the enveloping algebra are obtained from the symmetrization map (\ref{syma}). 

\medskip
\noindent It follows that the orders of the $1+\left[\frac{p+q}{2}\right]$ non-central invariants of $\mathfrak{Gal}_{\ell}(p,q)$ are 
\begin{itemize}
\item $4,4,8,\cdots ,2(p+q-1)$ if $d=p+q$ is odd, 

\item $4,4,8,\cdots ,2(p+q)-4,p+q$ if $d=p+q$ is even. 
\end{itemize}

\section{The unextended case}

\noindent As the centre of the Lie algebra $\mathfrak{Gal}_{\ell}(p,q)$ is one-dmensional, the corresponding factor algebra  $\overline{\mathfrak{Gal}}_{\ell}(p,q)=\mathfrak{Gal}_{\ell}(p,q)/Z\left(\mathfrak{Gal}_{\ell}(p,q)\right)$ inherits the structure of a semidirect product of the semisimple Lie algebra $\mathfrak{sl}(2,\mathbb{R})\oplus \mathfrak{so}(p,q)$ with the Abelian Lie algebra of dimension $d(2\ell+1)$, where the characteristic representation $\Gamma$ is given by $D_{\ell}\otimes \rho_1$. As this Lie algebra contains in particular the affine Lie algebra $\mathfrak{sl}(2,\mathbb{R})\overrightarrow{\oplus}_{D_{\ell}^{d}} \mathbb{R}^{(2\ell d+d)}$ as well as the multiply-inhomogeneous algebra $\mathfrak{so}(p,q)\overrightarrow{\oplus}_{\rho^{2\ell+1}} \mathbb{R}^{(2\ell d+d)}$, it is expected that the number of Casimir invariants of $\overline{\mathfrak{Gal}}_{\ell}(p,q)$ will be much higher than that of $\mathfrak{Gal}_{\ell}(p,q)$. An exception is given by the special case $\overline{\mathfrak{Gal}}_{\frac{1}{2}}(p,q)$, isomorphic to the unextended Schr\"odinger algebra $\widehat{\mathcal{S}}(p+q)$, for which the number of invariants is given by $\mathcal{N}(\widehat{\mathcal{S}}(p+q))=1+\left[\frac{p+q}{2}\right]$, constituting the only case where the number of (non-central) Casimir operators of the extension is preserved when passing to the factor Lie algebra.

\begin{proposition}\label{pro4}
For any $\ell\in \mathbb{Z}+\frac{1}{2}\geq \frac{1}{2}$ and $p+q=d\geq 3$ the number $\mathcal{N}(\mathfrak{g})$ of Casimir operators of $\overline{\mathfrak{Gal}}_{\ell}(p,q)$ is given by 
\begin{eqnarray}
\mathcal{N}(\mathfrak{g})=\left\{
\begin{array}[c]{rc}
1+\left[\frac{d}{2}\right], & \ell=\frac{1}{2},\; d\geq 3\\[0.1cm]
\frac{1}{2}\left(4\ell d+3d-d^2-6\right), & \ell\geq \frac{3}{2},\;  d\leq 2\ell+2\\[0.1cm]
2\ell^2+2\ell-\frac{5}{2}+\left[\frac{d}{2}\right], &  \ell\geq \frac{3}{2},\; d\geq 2\ell+3 \\
\end{array}
\right.
\end{eqnarray}
\end{proposition}

\noindent To prove the assertion, the best strategy is to use the reformulation of the formula (\ref{BB}) in terms of differential forms \cite{C43}. 
Let $\left\{\theta_1,\theta_2,\theta_3,\omega_{i,j},\sigma_{n,j}\right\}$ with $1\leq i,j\leq d$, $0\leq n\leq 2\ell$ be a basis of 1-forms dual to the basis $\left\{H,D,C,E_{i,j},P_{n,j}\right\}$ of  $\overline{\mathfrak{Gal}}_{\ell}(p,q)$. Then the Maurer-Cartan equations are given by 

\begin{eqnarray}
d\theta_1=-\theta_2\wedge\theta_3,\quad d\theta_2=2\theta_1\wedge\theta_2,\quad d\theta_3=-2\theta_1\wedge\theta_3,\nonumber\\
d\omega_{i,j}=\sum_{s=1}^{d} g_{ss} \omega_{i,s}\wedge\omega_{j,s},\quad 1\leq i<j\leq d,\nonumber\\
d\sigma_{0,j}=2\ell \theta_1\wedge\sigma_{0,j}-\theta_2\wedge\sigma_{1,j}+\sum_{s=1}^{d}g_{ss}\omega_{s,j}\wedge\sigma_{0,s},,\quad 1\leq j\leq d,\label{MCA}\\
d\sigma_{n,j}=2(\ell-n) \theta_1\wedge\sigma_{n,j}-(n+1)\theta_2\wedge\sigma_{n+1,j}+(2\ell+1-n)\theta_3\wedge\sigma_{n-1,j}\nonumber\\
\quad +\sum_{s=1}^{d}g_{ss}\omega_{s,j}\wedge\sigma_{n,s},\quad 1\leq n\leq 2\ell-1,\; 1\leq j\leq d,\nonumber\\
d\sigma_{2\ell,j}=-2\ell \theta_1\wedge\sigma_{2\ell,j}+\theta_3\wedge\sigma_{2\ell-1,j}+\sum_{s=1}^{d}g_{ss}\omega_{s,j}\wedge\sigma_{n,s},,\quad 1\leq j\leq d,.\nonumber 
\end{eqnarray}
We first consider the case $\ell=\frac{1}{2}$ corresponding to the unextended Schr\"{o}dinger algebra. For $%
d\leq 4$ the assertion follows at once considering the 2-form 
\begin{equation*}
\Xi _{1}=d\sigma _{0,1}+d\sigma _{1,d},
\end{equation*}
that has rank $5$ for $d=3$ and rank $7$ for $d=4$ respectively. For values $%
d\geq 5$ we define the forms 
\begin{equation*}
\Xi _{1}=d\sigma _{0,1}+d\sigma _{1,d},\;\Xi _{2}=\sum_{s=0}^{\alpha
}d\omega _{2+2s,3+2s},\;\alpha =\frac{2d-11-\left( -1\right) ^{d}}{4}.
\end{equation*}
Proceeding by induction, it can be easily shown that the product 
\begin{equation*}
\bigwedge^{d+1}d\sigma _{0,1}\bigwedge^{d-2}d\sigma
_{0,1}\bigwedge^{d-4}d\omega _{2,3}\cdots \bigwedge^{d-4-2\alpha }d\sigma
_{2+2\alpha ,3+2\alpha }
\end{equation*}
contains all of the 1-forms associated to generators of the Lie algebra $
\widehat{\mathcal{S}}\left( d\right) $ with the following exceptions 
\begin{equation}
\theta _{3},\omega _{2,3},\omega _{4,5},\cdots ,\omega _{d-2,d-1},\sigma
_{1,d}.  \label{exe}
\end{equation}
Counting the latter elements we conclude that 
\begin{equation}
2d-1+\sum_{s=0}^{\alpha }\left( d-4-2s\right) =\mu =\frac{1}{4}\left(
d^{2}+3d-4-2\left[ \frac{d}{2}\right] \right) .  \label{exec}
\end{equation}
Therefore, taking the 2-form $\Xi =\Xi _{1}+\Xi _{2}$, it is straightforward
to verify that it satisfies 
\begin{equation*}
\bigwedge^{\mu }\Xi =\bigwedge^{d+1}d\sigma _{0,1}\bigwedge^{d-2}d\sigma
_{0,1}\bigwedge^{d-4}d\omega _{2,3}\cdots \bigwedge^{d-4-2\alpha }d\omega
_{2+2\alpha ,3+2\alpha }+\cdots \neq 0,
\end{equation*}
showing that 
\begin{equation*}
\mathcal{N}\left( \widehat{\mathcal{S}}\left( d\right) \right) =1+\left[ \frac{d}{2}\right] .
\end{equation*}

\noindent This argumentation, with slight modifications, generalizes naturally for any value $\ell\geq \frac{3}{2}$, where it is also necessary to distinguish two cases,  depending whether $d=p+q\leq 2\ell +2$ or $d>2\ell+2$.

\begin{enumerate}
\item Let $d=p+q\leq 2\ell +2$. In this case the dimension of the characteristic representation $\Gamma$ is clearly larger than that of the Levi subalgebra, so that a 2-form of maximal rank can be constructed using only the differential forms associated to the generators $P_{n,k}$. Consider the 2-form in (\ref{MCA}) given by $\Theta=\Theta_1+\Theta_2$, where 
\begin{eqnarray}
\Theta_1=d\sigma_{0,1}+d\sigma_{2\ell,d}+d\sigma_{2\ell-1,d-1},\; 
\Theta_2=\sum_{s=1}^{d-4} d\sigma_{s,s+1}.\label{difo1}
\end{eqnarray}
Using the decomposition formula $\bigwedge^{a}\Theta=\sum_{r=0}^{a} \left(\bigwedge^{r}\Theta_1\right) \wedge \left(\bigwedge^{a-r}\Theta_2\right)$ we obtain that 
\begin{eqnarray}
\fl \bigwedge^{\frac{1}{2}\left(6-d+d^2\right)}\Theta= &\bigwedge^{d+1}d\sigma_{0,1}\wedge\bigwedge^{d-1}d\sigma_{2\ell,d}\wedge\bigwedge^{d-3}d\sigma_{2\ell-1,d-1}\wedge
\bigwedge^{d-4}d\sigma_{1,2}\wedge\nonumber\\
& \wedge\bigwedge^{d-5}d\sigma_{2,3}\wedge\bigwedge^{d-6}d\sigma_{3,4}\wedge\cdots \bigwedge^{2}d\sigma_{d-5,d-4}\wedge d\sigma_{d-4,d-3}+\cdots \neq 0.\label{pro2}
\end{eqnarray}
As $\frac{1}{2}\left(6-d+d^2\right)=\dim\left(\mathfrak{sl}(2,\mathbb{R})\oplus\mathfrak{so}(p,q)\right)$, the 2-form $\Theta$ is necessarily of maximal rank, as all the generators of the Levi subalgebra appear in some term of the product (\ref{pro2}) and no products of higher rank are possible due to the Abelian nilradical. We therefore conclude that $j(\mathfrak{g})=\frac{1}{2}\left(6-d+d^2\right)$ and by formula (\ref{BB1}) we have 
\begin{equation}
\mathcal{N}(\mathfrak{g})= \frac{1}{2}\left(4\ell d+3d-d^2-6\right).\label{inva1}
\end{equation}

\item Now let $d \geq 2\ell +3$. The main difference with respect to the previous case is that a generic form $\omega\in\mathcal{L}(\mathfrak{g})$ of maximal rank must necessarily contain linear combinations of the 2-forms $d\omega_{i,j}$ corresponding to the semisimple part of $\overline{\mathfrak{Gal}}_{\ell}(p,q)$. Let us consider first  the 2-form 
\begin{equation}
\Xi_1= \Theta_1+\Theta_2,
\end{equation}
where $\Theta_1$ is the same as in (\ref{difo1}) and $\Theta_2$ is defined as
\begin{equation}
\Theta_2=\sum_{s=0}^{2\ell-3} d\sigma_{1+s,2+s}.
\end{equation}
In analogy with the previous case, for the index  $\mu_1=(2\ell+1)d+(\ell+2)(1-2\ell)$ the first term of the following product does not vanish: 
\begin{equation}
\fl \bigwedge^{\mu_1}\Xi_1=\bigwedge^{d+1}d\sigma_{0,1}\bigwedge^{d-1}d\sigma_{2\ell,d}\bigwedge^{d-3}d\sigma_{2\ell-1,d-1} 
\bigwedge^{d-4}d\sigma_{1,2}\cdots \bigwedge^{d-1-2\ell}d\sigma_{2\ell-2,2\ell-1}+\cdots \neq 0.\label{Pot1}
\end{equation}
This form, although not maximal in $\mathcal{L}(\mathfrak{g})$, is indeed of maximal rank when restricted to the subspace $\mathcal{L}(\mathfrak{r})$ generated by the 2-forms $d\sigma_{n,k}$ with $0\leq n\leq 2\ell$, $1\leq k\leq d$. 
This means that the wedge product of $\bigwedge^{\mu_1}\Xi_1$  with any other $d\sigma_{n,k}$ is identically zero. Hence, in order to construct a 2-form of maximal rank in $\mathcal{L}(\mathfrak{g})$, we have to consider a 2-form $\Xi_2$ that is a linear combination of the  differential forms associated to the generators of the Levi subalgebra of $\overline{\mathfrak{Gal}}_{\ell}(p,q)$. As follows at once from (\ref{Pot1}), the forms $\theta_1,\theta_2,\theta_3$ associated to $\mathfrak{sl}(2,\mathbb{R})$-generators have already appeared, thus it suffices to restrict our analysis to linear combinations of the forms $d\omega_{i,j}$ corresponding to the pseudo-orthogonal Lie algebra $\mathfrak{so}(p,q)$. Specifically, we make the choice 
\begin{equation}
\Xi_2= \sum_{s=0}^{\nu}d\omega_{3+2s,4+2s},\quad \nu=\frac{1}{4}\left(2d-4\ell-9+(-1)^{1+d}\right).
\end{equation} 
Consider the integer $\mu_2=\frac{1}{4}\left(11+(d-4\ell)(1+d)-4\ell^2-2\left[\frac{d}{2}\right]\right)$ and take the 2-form $\Xi=\Xi_1+\Xi_2$. A long but routine computation shows that following identity is satisfied:
\begin{eqnarray}
\fl \bigwedge^{\mu_1+\mu_2}\Xi =& \left(\bigwedge^{\mu_1}\Xi_1\right)\wedge \left(\bigwedge^{\mu_2}\Xi_2\right) \nonumber\\
& =  \left(\bigwedge^{\mu_1}\Xi_1\right)\wedge\bigwedge^{d-6}d\omega_{3,4}\bigwedge^{d-8}d\omega_{5,6}\cdots \bigwedge^{d-6-2\nu}d\omega_{3+2\nu,4+2\nu}+\cdots \neq 0.\label{pro1}
\end{eqnarray}
We observe that this form involves $\mu_1+2\mu_2$ forms $\omega_{i,j}$ from $\mathfrak{so}(p,q)$, hence there remain $\frac{d(d-1)}{2}-\mu_1-2\mu_2$ elements of the pseudo-orthogonal that do not appear in the first term in (\ref{pro1}). From this product and (\ref{MCA}) it can be seen that these uncovered elements are of the type $\left\{\omega_{i_1,i_1+1},\omega_{i_2,i_2+1},\cdots \omega_{i_r,i_r+1}\right\}$ with the subindices satisfying $i_{\alpha+1}-i_{\alpha}\geq 2$ for $1\leq \alpha\leq r$, from which we deduce that no other 2-form $d\omega_{i_\alpha,i_\alpha+1}$, when multiplied with  $\bigwedge^{\mu_1+\mu_2}\Xi $ will be different from zero. 
We conclude that $\Xi$ has maximal rank equal to  $j_0(\mathfrak{g})=\mu_1+\mu_2$, thus applying (\ref{BB1}) we find that 
\begin{equation}
\fl \mathcal{N}(\mathfrak{g})= 3 + \frac{d(d-1)}{2}+ (2 \ell + 1) d-2(\mu_1+\mu_2)=  2\ell^2+2\ell-\frac{5}{2}+\left[\frac{d}{2}\right],
\end{equation}
as asserted.
\end{enumerate}

\medskip
\noindent In Table \ref{Tabelle1} we give the numerical values for the number of Casimir operators of the Lie algebras $\overline{\mathfrak{Gal}}_{\ell}(p,q)$ with $d=p+q\leq 12$, and where the linear increment with respect to $\ell$ can be easily recognized. 
 
\smallskip
\begin{table}[h!] 
\caption{\label{Tabelle1} Number of Casimir operators for $\overline{\mathfrak{Gal}}_{\ell}(p,q)$.}
\begin{indented}\item[]
\begin{tabular}{c||cccccccccc}
$\;d$ & $3$ & $4$ & $5$ & $6$ & $7$ & $8$ & $9$ & $10$ & $11$ & $12$ \\\hline 
{$\ell=\frac{1}{2}$} & $2$ & $3$ & $3$ & $4$ & $4$ & $5$ & $5$
& $6$ & $6$ & $7$ \\ 
{$\ell=\frac{3}{2}$} & $6$ & $7$ & $7$ & $8$ & $8$ & $9$ & $9$
& $10$ & $10$ & $11$ \\ 
{$\ell=\frac{5}{2}$} & $12$ & $15$ & $17$ & $18$ & $18$ & $19$
& $19$ & $20$ & $20$ & $21$ \\ 
{$\ell=\frac{7}{2}$} & $18$ & $23$ & $27$ & $30$ & $32$ & $33$
& $33$ & $34$ & $34$ & $35$ \\ 
{$\ell=\frac{9}{2}$} & $24$ & $31$ & $37$ & $42$ & $46$ & $49$
& $51$ & $52$ & $52$ & $53$ \\ 
{$\ell=\frac{11}{2}$} & $30$ & $39$ & $47$ & $54$ & $60$ & $65
$ & $69$ & $72$ & $74$ & $75$%
\end{tabular}
\end{indented}
\end{table}

\medskip
\noindent As follows from a general property concerning virtual copies \cite{C45}, Lie algebras of the type $\mathfrak{g}=\mathfrak{s}\overrightarrow{\oplus} \mathfrak{r}$ with an Abelian radical $\mathfrak{r}$ do not admit virtual copies of $\mathfrak{s}$ in $\mathcal{U}\left(\mathfrak{g}\right)$. Thus for Lie algebras of this type the Casimir invariants must be computed either directly from system (\ref{sys}) or by some other procedure. Among the class  $\overline{\mathfrak{Gal}}_{\ell}(p,q)$, an exception is given by the unextended (pseudo-)Schr\"odinger algebra  $\overline{\mathfrak{Gal}}_{\frac{1}{2}}(p,q)\simeq \widehat{\mathcal{S}}(p,q)$, where the invariants can be deduced from those of the central extension $\widehat{\mathcal{S}}(p,q)$ by the widely used method of contractions (see e.g. \cite{IW,We}). For the remaining values $\ell\geq \frac{3}{2}$ the contraction procedure is useless in practice, given the high number of invariants.  However, an interesting property concerning the invariants of $\overline{\mathfrak{Gal}}_{\ell}(p,q)$ emerges when we try to find the Casimir operators $F$ that only depend on variables $p_{n,k}$ associated to generators $P_{n,k}$ of the radical, i.e., such that the condition 
\begin{equation}
\quad \frac{\partial F}{\partial x}=0,\quad \forall x\in\mathfrak{sl}(2,\mathbb{R})\oplus\mathfrak{so}(p,q).\label{kond}
\end{equation}
is satisfied. As will be shown next, the number of such solutions tends to stabilize for high values of $d=p+q$, showing that almost any invariant will depend on all of the variables in $\overline{\mathfrak{Gal}}_{\ell}(p,q)$, implying that finding a complete set of invariants is a computationally formidable task, as there is currently no general method to derive these invariants in closed form. 

\begin{proposition}
Let $\ell\geq \frac{3}{2}$. For sufficiently large $d$, the number of Casimir invariants of  $\overline{\mathfrak{Gal}}_{\ell}(p,q)$ depending only on the variables $p_{n,k}$ of the Abelian radical is constant and given by 
\begin{equation}
\mathcal{N}_1(S)=2\ell^2+3\ell-2.\label{sr2}
\end{equation}
\end{proposition}

\noindent The proof follows analyzing the rank of the subsystem of (\ref{sys}) corresponding to the differential operators $\widehat{X}$ associated to the generators of the Levi subalgebra $\mathfrak{sl}(2,\mathbb{R})\oplus\mathfrak{so}(p,q)$ and such that condition (\ref{kond}) is fulfilled. Specifically, this leads to the system $S$ of PDEs
\begin{eqnarray}
\widehat{D}^{\prime}(F):=\sum_{n=0}^{2\ell}\sum_{i=1}^{d} (2\ell-n)p_{n,i}\frac{\partial F}{\partial p_{n,i}}=0,\; 
\widehat{H}^{\prime}(F):=\sum_{n=0}^{2\ell}\sum_{i=1}^{d} n p_{n-1,i}\frac{\partial F}{\partial p_{n,i}}=0,\nonumber\\
\widehat{C}^{\prime}(F):=\sum_{n=0}^{2\ell}\sum_{i=1}^{d} (2\ell-n)p_{n+1,i}\frac{\partial F}{\partial p_{n,i}}=0,\label{kond2}\\
\widehat{E}_{j,k}^{\prime}(F):=\sum_{n=0}^{2\ell}\sum_{i=1}^{d} \left( g_{ij} p_{n,k} -g_{ik} p_{n,j}\right) \frac{\partial F}{\partial p_{n,i}}=0, 1\leq j<k\leq d.\nonumber
\end{eqnarray}
This system consists of $\frac{1}{2}\left(6-d+d^2\right)$ equations in $(2\ell+1)d$ variables that becomes overdetermined for increasing values of $d$ (and fixed $\ell$). In Table \ref{Tabelle2} the rank of such systems is given for values $d\leq 15$, showing that for fixed $\ell$, from $d\geq 2\ell+1$ onwards, the rank of the system increases always by the same constant amount, given precisely by $2\ell+1$.  

\begin{table}[h!] 
\caption{\label{Tabelle2} Rank of system (\ref{kond2}).}
\begin{indented}\item[]
\begin{tabular}{c||ccccccccccccc}
$d$ & $3$ & $4$ & $5$ & $6$ & $7$ & $8$ & $9$ & $10$ & $11$ & $12$ & $13$& $14$& $15$ \\ \hline
$\ell =\frac{3}{2}$ & 6 & 9 & 13 & 17 & 21 & 25 & 29 & 33 & 37 & 41 & 45 & 49  & 53\\
$\ell =\frac{5}{2}$ & 6 & 9 & 13 & 18 & 24 & 30 & 36 & 42 & 48 & 54 & 60 & 66 & 72\\
$\ell =\frac{7}{2}$ & 6 & 9 & 13 & 18 & 24 & 31 & 39 & 47 & 55 & 63 & 71 & 79
& 87 \\ 
$\ell =\frac{9}{2}$ & 6 & 9 & 13 & 18 & 24 & 31 & 39 & 48 & 58 & 68 & 78 & 88
& 98 \\ 
$\ell =\frac{11}{2}$ & 6 & 9 & 13 & 18 & 24 & 31 & 39 & 48 & 58 & 69 & 81 &
93 & 105 \\ 
$\ell =\frac{13}{2}$ & 6 & 9 & 13 & 18 & 24 & 31 & 39 & 48 & 58 & 69 & 81 & 94 & 108
\end{tabular}
\end{indented}
\end{table}
\noindent With these observations, it is not difficult to establish that for any $\ell\geq \frac{3}{2}$ and $d\geq 2\ell+1$ the rank of the system (\ref{kond2}) is given by 
\begin{equation}
{\rm rank}\; S =\left(2+d\right)+\ell\left(2d-3\right)-2\ell^2.\label{kond3}
\end{equation}
As the number of variables is $(2\ell+1)d$, we conclude that the system admits exactly
\begin{equation}
\mathcal{N}_1(S)= (2\ell+1)d- {\rm rank}\; S = 2\ell^2+3\ell-2
\end{equation}
solutions satisfying the constraint (\ref{kond}). Further, comparison with Proposition \ref{pro4} allows us to establish that for any fixed $\ell$ and $d\leq 2\ell+2$, the following identity holds:
\begin{equation}
\mathcal{N}\left(\overline{\mathfrak{Gal}}_{\ell}(p,q)\right)=\mathcal{N}_1(S).\label{trox}
\end{equation}
For increasing values of $d$, there appear additional invariants that necessarily depend on variables associated to the generators of the Levi subalgebra of $\overline{\mathfrak{Gal}}_{\ell}(p,q) $. 

\medskip
\noindent Although there is currently no algorithmic procedure to construct a complete set of invariants of these Lie algebras for arbitrary values $d>2\ell+2$, those invariants of $\mathfrak{Gal}_{\ell}(p,q)$ satisfying the condition (\ref{kond}) can be easily computed by means of a reduction argument that leads to a linear system. To this extent, consider the last of the equations in (\ref{kond2}). As the generators of $\mathfrak{so}(p,q)$ permute the generators of the Abelian radical, it is straightforward to verify that the quadratic polynomials 
\begin{equation}
\Phi_{n,s}= \sum_{k=1}^{d} \frac{g_{11}}{g_{kk}}\;p_{n,k}p_{n+s,k},\; 0\leq n\leq 2\ell,\; 0\leq s\leq 2\ell-n.\label{ELE}
\end{equation}
are actually solutions of these equations. Indeed, any solution of the type (\ref{kond}) is built up from these functions. Let $\mathcal{M}_d=\left\{\Phi_{n,s},\; 0\leq n\leq 2\ell,\; 0\leq s\leq 2\ell-n\right\}$. The cardinal of this set is given by $2\ell^2+3\ell+1$, and we observe that not all of the elements in $\mathcal{M}_d$ are independent. It follows by a short computation that 
\begin{equation}
\widehat{D}^{\prime}(\mathcal{M}_d)\subset \mathcal{M}_d,\; \widehat{H}^{\prime}(\mathcal{M}_d)\subset \mathcal{M}_d,\; \widehat{C}^{\prime}(\mathcal{M}_d)\subset \mathcal{M}_d,\label{ELE2}
\end{equation}
showing that this set is invariant by the action of $\mathfrak{sl}(2,\mathbb{R})$. Therefore, we can construct the solutions of system (\ref{kond2}) recursively using polynomials in the new variables $\Phi_{n,s}$. Specifically, renumbering the elements in $\mathcal{M}_d$ as $\left\{u_{1},\cdots ,u_{2\ell^2+3\ell+1}\right\}$, for any $r\geq 2$ we define a polynomial of degree $2r$ as  
\begin{equation}
\Psi_r= \sum_{1\leq i_1< \cdots <i_r\leq |\mathcal{M}_d|} \alpha^{i_1\cdots i_r}  u_{i_1}u_{i_2}\cdots u_{i_r},\; i_1+\cdots i_r=r.\label{poly}
\end{equation}
Now, imposing the constraints
\begin{equation}
\widehat{D}^{\prime}(\Psi_r)=0,\; \widehat{H}^{\prime}(\Psi_r)=0,\; \widehat{C}^{\prime}(\Psi_r)=0,\label{ELE3}
\end{equation}
leads to a linear system in the coefficients $\alpha^{i_1\cdots i_r}$, the solutions of which enable us to find the polynomials that satisfy system (\ref{kond2}). Alternatively, the functions 
$\Phi_{n,s}$ can be used as new variables to reduce the equations in (\ref{ELE3}) to a simpler form, which may be computationally more effective, albeit the underlying argument is essentially the same \cite{Dick}. In the case where the identity (\ref{trox}) holds, this reduction procedure allows us to obtain a complete set of invariants for the Lie algebra $\overline{\mathfrak{Gal}}_{\ell}(p,q) $.

\medskip
\noindent As an example to illustrate the reduction, consider the 18-dimensional Lie
algebra $\overline{\frak{Gal}}_{\frac{3}{2}}\left( 3\right) $. As $d<2\ell+2$,
formula (\ref{trox}) applies and the algebra has 6 Casimir operators. From these,
two of order four in the generators can be derived from the central
extension $\frak{Gal}_{\frac{3}{2}}\left( 3\right) $ by contraction \cite
{We}. In this case, the set $\mathcal{M}_{3}$ has ten elements that we
enumerate as follows:      
\begin{equation*}
\left\{ \Phi _{00},\Phi _{01},\Phi _{02},\Phi _{03},\Phi _{10},\Phi
_{11},\Phi _{12},\Phi _{20},\Phi _{21},\Phi _{30}\right\} =\left\{
u_{1},\cdots ,u_{10}\right\} .
\end{equation*}
The action of the differential operators associated to $\frak{sl}\left( 2,%
\mathbb{R}\right) $ on $\mathcal{M}_{3}$ is explicitly given in Table \ref{Tabelle3}.

\begin{table}[h!] 
\caption{\label{Tabelle3} Transformation rules of variables $u_i$ under the $\mathfrak{sl}(2,\mathbb{R})$-action (\ref{kond2}).}
%\begin{indented}\item[]
\footnotesize\rm
\begin{tabular}{@{}*{1}{c|cccccccccc}}
& $u_{1}$ & $u_{2}$ & $u_{3}$ & $u_{4}$ & $u_{5}$ & $u_{6}$ & $u_{7}$ & $%
u_{8}$ & $u_{9}$ & $u_{10}$ \\[0.1cm] \hline  
$\widehat{D}^{\prime }$ & $6u_{1}$ & $4u_{2}$ & $2u_{3}$ & $0$ & $2u_{5}$ & $%
0$ & $-2u_{7}$ & $-2u_{8}$ & $-4u_{9}$ & $-6u_{10}$ \\ 
$\widehat{H}^{\prime }$ & $0$ & $-u_{1}$ & $-2u_{2}$ & $-3u_{3}$ & $-2u_{2}$
& $-u_{3}-2u_{5}$ & $-u_{4}-3u_{6}$ & $-4u_{6}$ & $-2u_{7}-3u_{8}$ & $-6u_{9}
$ \\ 
$\widehat{C}^{\prime }$ & $6u_{2}$ & $2u_{3}+3u_{5}$ & $u_{4}+3u_{6}$ & $%
3u_{7}$ & $4u_{6}$ & $u_{2}+2u_{8}$ & $2u_{9}$ & $2u_{9}$ & $u_{10}$ & $0$%
\end{tabular}
%\end{indented}
\end{table}
It follows from this action that polynomials $\Psi _{r}$ in the $u_{i}$ that satisfy the system (\ref
{ELE3}) are the solutions of the following system of linear first-order
partial differential equations:

{\footnotesize
\begin{equation}
\fl
\begin{tabular}{rr}
$6u_{1}\frac{\partial F}{\partial u_{1}}+4u_{2}\frac{\partial F}{\partial
u_{2}}+2u_{3}\frac{\partial F}{\partial u_{3}}+2u_{5}\frac{\partial F}{%
\partial u_{5}}-2u_{7}\frac{\partial F}{\partial u_{7}}-2u_{8}\frac{\partial
F}{\partial u_{8}}-4u_{9}\frac{\partial F}{\partial u_{9}}-6u_{10}\frac{%
\partial F}{\partial u_{10}}$ & $=0,$ \\ 
$-u_{1}\frac{\partial F}{\partial u_{2}}-2u_{2}\frac{\partial F}{\partial
u_{3}}-3u_{3}\frac{\partial F}{\partial u_{4}}-2u_{2}\frac{\partial F}{%
\partial u_{5}}-\left( u_{3}+2u_{5}\right) \frac{\partial F}{\partial u_{6}}%
-\left( u_{4}+3u_{6}\right) \frac{\partial F}{\partial u_{7}}-4u_{6}\frac{%
\partial F}{\partial u_{8}}$ &  \\ 
$-\left( 2u_{7}+3u_{8}\right) \frac{\partial F}{\partial u_{9}}-6u_{9}\frac{%
\partial F}{\partial u_{10}}$ & $=0,$ \\ 
$6u_{2}\frac{\partial F}{\partial u_{1}}+\left( 2u_{3}+3u_{5}\right) \frac{%
\partial F}{\partial u_{2}}+\left( u_{4}+3u_{6}\right) \frac{\partial F}{%
\partial u_{3}}+3u_{7}\frac{\partial F}{\partial u_{4}}+4u_{6}\frac{\partial
F}{\partial u_{5}}+\left( u_{2}+2u_{8}\right) \frac{\partial F}{\partial
u_{6}}+2u_{9}\frac{\partial F}{\partial u_{7}}$ &  \\ 
$+2u_{9}\frac{\partial F}{\partial u_{8}}+u_{10}\frac{\partial F}{\partial
u_{9}}$ & $=0.$%
\end{tabular}\label{reda}
\end{equation}
}
\noindent This system admits two quadratic solutions given by 
\begin{eqnarray*}
F_{1}
&=&3u_{4}^{2}+27u_{6}^{2}-18u_{3}u_{7}-27u_{5}u_{8}+12u_{2}u_{9}-3u_{1}u_{10},
\\
F_{2}
&=&27u_{6}^{2}-5u_{4}^{2}+18u_{4}u_{6}+12u_{3}u_{7}-4u_{1}u_{10}+24u_{2}u_{9}-36\left( u_{5}u_{7}+u_{3}u_{8}\right) .
\end{eqnarray*} 
Incidentally, these are the invariants that are obtained by contraction from those of the centrally-extended algebra $\mathfrak{Gal}_{
\frac{3}{2}}\left( 3\right) $. 
In addition, there exist four additional independent fourth-order solutions,
the explicit expression of which is omitted because of its length. We
conclude that a complete set of Casimir operators of $\overline{\frak{Gal}}_{%
\frac{3}{2}}\left( 3\right) $ is given by two fourth-order polynomials in
the generators (corresponding to the quadratic solutions of (\ref{reda}))
and four invariants of order eight corresponding to the fourth-order
solutions of (\ref{reda}).

\section{Final remarks}

We have seen that the generalized conformal Galilean algebras $\widehat{\mathfrak{g}}_{\ell}(d)$ based on the semisimple Lie algebra $\mathfrak{sl}(2,\mathbb{R})\oplus\mathfrak{so}(d)$ can be extended naturally to pseudo-Galilean algebras possessing a Levi subalgebra isomorphic to $\mathfrak{sl}(2,\mathbb{R})\oplus\mathfrak{so}(p,q)$ introducing a nondegenerate metric tensor into the orthogonal part. Virtual copies of $\mathfrak{sl}(2,\mathbb{R})\oplus\mathfrak{so}(p,q)$ in the enveloping algebra of the semidirect product can be obtained simultaneously for all (half-integer) values of $\ell$ and $p+q=d$. The resulting Lie algebras $\mathfrak{Gal}_{\ell}\left( p,q\right) $ can be seen, to a certain extent, as ``real" forms of the conformal Galilean algebra $\widehat{\mathfrak{g}}_{\ell}(d)$, their main structural difference residing in the maximal compact subalgebra. Whether these Lie algebras $\mathfrak{Gal}_{\ell}\left( p,q\right) $ have some definite physical meaning is still an unanswered question, but it is conceivable that they appear in the context of dynamical groups of higher order Lagrangian systems or as the (maximal) invariance symmetry group of a (hierachy of) partial differential equations. The search of physical realizations of the Lie algebras $\mathfrak{Gal}_{\ell}\left( p,q\right) $ is currently being developed.

\smallskip
\noindent We observe that the obstructions found for integer values of $\ell$ and leading to the so-called exotic extensions (see e.g. \cite{Als19} and reference therein) are a direct consequence of the incompatibility of the odd-dimensional representation $D_{\ell}$ with a Heisenberg algebra. Indeed, as shown in \cite{C45}, the necessary and sufficient condition for a semidirect product $\mathfrak{s}\overrightarrow {\oplus}_{\Gamma\oplus \Gamma_0}\mathfrak{h}_n$ to exist is that the (nontrivial) characteristic representation $\Gamma$ satisfies the condition $\Gamma\wedge \Gamma\supset \Gamma_0$. For the decomposition of $\Gamma$ into irreducible components, this implies in particular that an irreducible representation of $\mathfrak{s}$ must appear with the same multiplicity as its dual or be self-dual. Therefore, in order to further generalize the notion of Galilean algebras to $\mathfrak{sl}(2,\mathbb{R})$-representations with even highest weight, the characteristic representation $\Gamma$ must have the form
\begin{equation}
\Gamma =\left(D_\ell\oplus D_\ell\right)\otimes \rho_d
\end{equation}
As happens with any coupling of a semisimple Lie algebra $\mathfrak{s}$ and a Heisenberg Lie algebra $\mathfrak{h}_n$, the (noncentral) Casimir operators of the semidirect product 
$\mathfrak{s}\overrightarrow {\oplus}_{\Gamma\oplus \Gamma_0}\mathfrak{h}_n$ can be constructed using the invariants of  $\mathfrak{s}$ by means of the virtual copy method \cite{Que,C45}. Application of this procedure in combination with the trace method provides explicit expressions for the invariants of $\mathfrak{Gal}_{\ell}\left( p,q\right) $ for arbitrary values of $\ell$ and $p+q=d$, comprising in particular the case $\widehat{\mathfrak{g}}_{\ell}(d)=\mathfrak{Gal}_{\ell}\left( d,0\right) $ recently announced \cite{raub}. 

\medskip
\noindent  The case of the unextended conformal pseudo-Galilean algebra  $\overline{\mathfrak{Gal}}_{\ell}(p,q) $ corresponding to the factor of $\mathfrak{Gal}_{\ell}(p,q) $ by its centre has also been considered. As this Lie algebra has an Abelian radical, it does not admit a virtual copy in the corresponding enveloping algebra, hence their invariants must be computed by other means. The number of Casimir operators for arbitrary values of the parameters has been computed by means of the Maurer-Cartan equations of $\overline{\mathfrak{Gal}}_{\ell}(p,q)$, where a varying increasement behaviour for the number of invariants in dependence of the proportion between the dimension of the pseudo-orthogonal subalgebra and the dimension $2\ell+1$ of the $\mathfrak{sl}(2,\mathbb{R})$-representation $D_\ell$ has been observed. Although explicit formulae for  the Casimir invariants of $\overline{\mathfrak{Gal}}_{\ell}(p,q) $ with $\ell\geq \frac{3}{2}$ can probably not be found generically, it has been shown that the functions depending only on variables of the radical provide a complete set of invariants for the Lie algebra whenever the condition $d\leq 2\ell+2$ is satisfied. A procedure that reduces the computation of such invariants to solving a linear system has been proposed. However, even with this systematization, the problem still involves cumbersome computations, as the orders of such invariants are quite elevated and there is currently no result that allows to predict these orders. For values $d\geq 2\ell+3$, where there exist Casimir operators that do not satisfy the condition (\ref{kond}), no valuable ansatz has been found that allows to find them systematically. Any kind of progress in this direction would constitute a useful tool for the generic analysis of invariant functions of semidirect products of semisimple and Abelian Lie algebras, a class that up to certain relevant special cases has still not been exhaustively studied. 

\medskip
\noindent

\section*{Acknowledgment}
During the
preparation of this work, the RCS was financially supported by
the research project MTM2016-79422-P of the AEI/FEDER (EU). IM was supported
by the Australian Research Council Discovery Grant DP160101376 and Future Fellowship FT180100099.


\section*{References}
\begin{thebibliography}{9}

\bibitem{Cu} Cunningham E 1910 {\it Proc. London Math. Soc.} {\bf 8} 77 (1909)

\bibitem{Bat} Bateman H {\it Proc. London Math. Soc.} {\bf  8} 228 

\bibitem{Ful} Fulton T, Rohrlich F and Witten L 1962 {\it Rev. Mod. Phys.} {\bf 34} 442

\bibitem{Bar54}
Bargmann V 1954 {\it Ann. Math.} {\bf 59} 1

\bibitem{Hag} Hagen C R 1972 {\it Phys. Rev.} {\bf D5} 377

\bibitem{Hav} Havas P and Plebanski J 1978 {\it J. Math. Phys.} {\bf 19} 482
 
\bibitem{Zak} Andrzejewski K 2014 {\it Phys. Lett. B} {\bf 738} 405

\bibitem{Fig} Figueroa-O'Farrill J M 2019 \textit{J. Math. Phys.} {\bf 60} 021702

\bibitem{Ni72}
Niederer U 1972 {\it Helv. Phys. Acta} {\bf 45} 802

\bibitem{Ni73}
Niederer U 1973 {\it Helv. Phys. Acta} {\bf 46} 191

\bibitem{Do97}
Dobrev V K, Doebner H D and Mrugalla C 1997 {\it Rep. Math. Phys.}
{\bf 39} 201

\bibitem{Fra} Campoamor-Stursberg R 2005 {\it J. Phys. A: Math. Gen.} \textbf{38}
4187

\bibitem{Ai12}
Aizawa N, Isaac P S and Kimura Y 2012 {\it Int. J. Math.} {\bf 23}
1250118

\bibitem{Tac} Martelli D and Tachikawa Y 2010 {\it JHEP} {\bf 5} 1

\bibitem{Du11}
Duval C and Horvathy P A 2011 {\it J. Phys. A: Math. Theor.} {\bf
44} 335

\bibitem{St13}
Stichel P C and Zakrzewski W J 2013 {\it Entropy} {\bf 15} 559

\bibitem{Als17}
Alshammari F, Isaac P S and Marquette I 2018 \textit{J. Phys. A:
Math. Theor.} {\bf 51} 065206

\bibitem{Als19}
Alshammari F, Isaac P S and Marquette I 2019 \textit{J. Math.
Phys.} {\bf 60} 013509

\bibitem{raub} Galajinsky A and Masterov I 2019 arXiv:1902.08012

\bibitem{Que} Quesne Ch 1988 {\it J. Phys. A: Math. Gen.} \textbf{21} L321

\bibitem{C23} Campoamor-Stursberg R 2004 {\it Acta Physica Pol. B} \textbf{35}  2059

\bibitem{C45} Campoamor-Stursberg R 2005  \textit{Acta Physica Pol. B} \textbf{36} (2005),
2869

\bibitem{SL3} Campoamor-Stursberg R and Low S G 2009 {\it J. Phys. A: Math. Gen.} \textbf{42} 065205

\bibitem{Tits} Tits J 1967 {\it Tabellen zu den einfachen Lie Gruppen und ihren Darstellungen} 
(Springer Verlag, Berlin)

\bibitem{Ra} Racah G 1950  {\it Rend. Sci. Fis. Mat. Nat.} {\bf 8} 108

\bibitem{Gel} Gel'fand I M 1950 \textit{Mat. Sb.} \textbf{50} 103

\bibitem{Bo1} Boyko V, Patera J and Popovych R 2006
{\it J. Phys. A: Math. Gen.} \textbf{39} 5749

\bibitem{Bo2} Boyko V, Patera J and Popovych R  2007 \textit{J. Phys. A: Math.
Theor.} \textbf{40} 113

\bibitem{Be} Beltrametti E G and Blasi A 1966 {\it Phys. Letters} {\bf 20}
62

\bibitem{C43} Campoamor-Stursberg R 2004 {\it Phys. Letters} \textbf{A327}
138

\bibitem{Peci} Pecina-Cruz J N 2006  {\it J. Math. Phys.} {\bf 46} 063503 

\bibitem{C72} Campoamor-Stursberg R 2008  \textit{J. Phys. A: Math. Theor.} {\bf 41} 365207

\bibitem{Gr64}
Gruber B, and O'Raifeartaigh L, {\it J. Math. Phys.} 5 (1964)
1796

\bibitem{Per}
Perelomov A M and Popov V S 1968 \textit{Izv. Akad. Nauk}
\textbf{32} 1368

\bibitem{Po66}
Popov V S 1977 {\it Teor. Mat. Fiz.} {\bf 32} 344.

\bibitem{Ok77}
Okubo S 1977 {\it J. Math. Phys.} {\bf 18} 2382

\bibitem{Mac} Macfarlane A J and Pfeiffer H 2001 {\it J. Math. phys.} \textbf{41} 3192

\bibitem{IW}  Saletan E 1961 {\it J. Math. Phys.} \textbf{2} 1

\bibitem{We} Weimar-Woods E 2008 {\it J. Math. Phys.} {\bf 49}
033507

\bibitem{Dick} Dickson L E 1924 \textit{Ann. Math.} {\bf 25} 287
 

\end{thebibliography}
\end{document}